\documentclass[twocolumn,notitlepage,prl,superscriptaddress,showpacs]{revtex4-1}
\usepackage{hyperref}
\usepackage[usenames,dvipsnames]{color}
\usepackage{amsmath}
\usepackage[english]{babel}
\usepackage{amssymb}
\usepackage{graphicx}
\usepackage{epstopdf}
\usepackage{tabularx}
\usepackage{multirow}
\usepackage[]{todonotes}

\begin{document}
%\linenumbers
\title{Extended Crossover from a Fermi Liquid to a Quasiantiferromagnet in the Half-Filled 2D Hubbard Model}

\author{Fedor \v{S}imkovic IV} \altaffiliation[Present address: ]{Centre de Physique Th\'eorique, \'Ecole Polytechnique,
CNRS, Universit\'e Paris-Saclay, 91128 Palaiseau, France and Coll\`{e}ge de France, 11 place Marcelin Berthelot, 75005 Paris, France.}

\affiliation{Department of Physics, King’s College London, Strand, London WC2R 2LS, UK}
\author{J. P. F. LeBlanc}
\affiliation{Department of Physics and Physical Oceanography, Memorial University of Newfoundland, St. John's, Newfoundland and Labrador A1B 3X7, Canada}
\author{Aaram J. Kim}
\affiliation{Department of Physics, King’s College London, Strand, London WC2R 2LS, UK}
\author{Youjin Deng}
\affiliation{Hefei National Laboratory for Physical Sciences at Microscale and Department of Modern Physics, University of Science and Technology of China, Hefei, Anhui 230026, China}
\author{N. V. Prokof'ev}
\affiliation{Department of Physics, University of Massachusetts, Amherst, Massachusetts 01003, USA}
\affiliation{National Research Center ``Kurchatov Institute," 123182 Moscow, Russia }
\author {B.V. Svistunov}
\affiliation{Department of Physics, University of Massachusetts, Amherst, Massachusetts 01003, USA}
\affiliation{National Research Center ``Kurchatov Institute," 123182 Moscow, Russia }
\affiliation{Wilczek Quantum Center, School of Physics and Astronomy and T. D. Lee Institute, Shanghai Jiao Tong University, Shanghai 200240, China}
\author{Evgeny Kozik}
\email{evgeny.kozik@kcl.ac.uk}
\affiliation{Department of Physics, King’s College London, Strand, London WC2R 2LS, UK}

\date{\today}

\begin{abstract}
The ground state of the Hubbard model with nearest-neighbor hopping on the square lattice at half filling is known to be that of an antiferromagnetic (AFM) band insulator for any on-site repulsion. At finite temperature, the absence of long-range order makes the question of how the interaction-driven insulator is realized nontrivial. We address this problem with controlled accuracy in the thermodynamic limit using self-energy diagrammatic determinant Monte Carlo and dynamical cluster approximation methods and show that development of long-range AFM correlations drives an extended crossover
from Fermi liquid to insulating behavior in the parameter regime that precludes a metal-to-insulator transition. The intermediate crossover state is best described as a non-Fermi liquid with a partially gapped Fermi surface.
\end{abstract}

\maketitle
%\listoftodos

%\section{Introduction}

The interaction driven metal-to-insulator (MIT) transition has been for many years a problem of central focus for the field of strongly correlated electron systems (see, e.g., Refs.~\cite{georges:1996, schafer:2015} and references therein).
Particularly challenging has been the quantitative, and even qualitative, understanding of the MIT in two-dimensional systems. Here, the basic model---the single-band Hubbard model with nearest-neighbor hopping on the square lattice---can nowadays be accurately emulated and probed with ultracold atoms in optical lattices \cite{Bloch_review_2005, lewenstein2007ultracold, greif2015formation, parsons2016site, greiner2017, nichols2018spin} at  ever decreasing temperatures, putting controlled experimental studies of the this problem within reach. The model is given by the Hamiltonian
\begin{equation}
H = \sum_{\mathbf{k},\sigma} \left(\epsilon_\mathbf{k} -\mu\right)c_{\mathbf{k}\sigma}^\dagger c_{\mathbf{k}\sigma}+U\sum_i n_{i\uparrow}n_{i\downarrow},
\label{H}
\end{equation}
where $\mu$ is the chemical potential, $\mathbf{k}$ (quasi)momentum with the lattice constant set to unity, $n_{i\sigma}$ the number operator of fermions with spin $\sigma$ on the square lattice site $i$, $U$ the on-site repulsion, and
$\epsilon_\mathbf{k}=-2t\left[\cos(k_x)+\cos(k_y)\right]$. For the description of the MIT, nonperturbative numerical methods, such as the dynamical mean-field theory (DMFT) and related cluster and diagrammatic extensions \cite{georges:1996, Maier05, park:2008, Werner098site, gull:2013, schafer:2015, rohringer2016, fratino2017afm, vanloon:2018:SO}, have played a central role. In the single-site paramagnetic DMFT \cite{georges:1996} solution, which becomes exact in the limit of infinite dimensions, the metallic phase at half filling (the average density per site $\langle n \rangle =1$)
persists down to zero temperature at weak interactions.
It is separated from the Mott insulator by a first-order MIT at a sufficiently large value of $U=U_c$, ending at a finite temperature with an Ising critical point. Extensions of DMFT to small (up to 16 sites) real-space clusters \cite{park:2008, Werner098site, gull:2013} have shown that the inclusion of short-range spin fluctuations changes this picture substantially---a non-Fermi-liquid (nFL) state with a Fermi surface (FS) gap in certain momentum sectors continuously develops at a finite $U$ before the transition, the value of $U_c$ is reduced, and the slope of the first-order line is inverted.   
%Nevertheless, these studies have not questioned the qualitative picture of the MIT for the model (\ref{H}).

It is, however, well known that the ground state of the model (\ref{H}) at $\langle n \rangle =1$
is an antiferromagnet (AFM) at any $U>0$.
As revealed by Slater~\cite{Slater:1951}, the FS nesting, i.e. the existence of a single wave vector $\mathbf{Q}=(\pi,\pi)$ that connects any point on the FS to another FS point, makes the interacting Fermi gas
unstable against formation of the spin density wave with the wave vector $\mathbf{Q}$ already at infinitesimally small $U$.
The corresponding unit-cell doubling makes the ground state a band insulator. While the Mermin Wagner theorem~\cite{Mermin-Wagner} forbids the long-range order at $T>0$, the AFM correlation length $\xi$
is exponentially large at low temperature, $\log\xi \propto t/T$; i.e., for practical purposes the system is best described
as a quasi-AFM. Indeed, experiments with ultracold atoms~\cite{greiner2017} observed a perfect
AFM state in model (\ref{H}) on a $\sim 10 \times 10$ lattice at temperatures as high as $T\sim 0.25t$.
Such AFM correlations are explicitly truncated (and typically suppressed) in cluster DMFT calculations unless
linear cluster sizes are comparable to $\xi$~\cite{leblanc:2013}, which becomes computationally prohibitive at low $T$ even at half filling~\cite{benchmarks}.
Recent work~\cite{schafer:2015} based on the dynamical vertex approximation (D$\Gamma$A), a diagrammatic extension of DMFT
capable of capturing long-range correlations approximately, and determinant quantum Monte Carlo (DQMC) simulations indicates
that the low-temperature crossover from the Fermi liquid (FL) to the quasi-AFM insulator preempts and precludes the MIT.

In this Letter, we aim at establishing the picture of developing an insulating AFM state
with controlled accuracy using the recently introduced $\Sigma$DDMC approach~\cite{simkovic2017determinant} (a similar approach was developed in Ref.~\cite{moutenet2018determinant}).
The method deterministically sums all topologies of Feynman diagrams for self-energy (for introduction, see, e.g., Ref.~\cite{AGD}) by means of determinants \cite{rubtsov2003, rubtsov2005continuous} with a recursive scheme in the spirit of Rossi's algorithm~\cite{Rossi2016det} to extract only one-particle irreducible diagrams. Integration over
internal variables is performed by Monte Carlo sampling, and the thermodynamic limit (TDL) is taken explicitly.
Compared to the standard DiagMC approach \cite{van2010diagrammatic, kozik2010diagrammatic}, where diagram topologies
are sampled stochastically, $\Sigma$DDMC enables access to substantially higher ($\sim 10-12$) expansion orders and more accurate determination of the self-energy.

\begin{figure}
\centering
\includegraphics[width=0.9\linewidth]{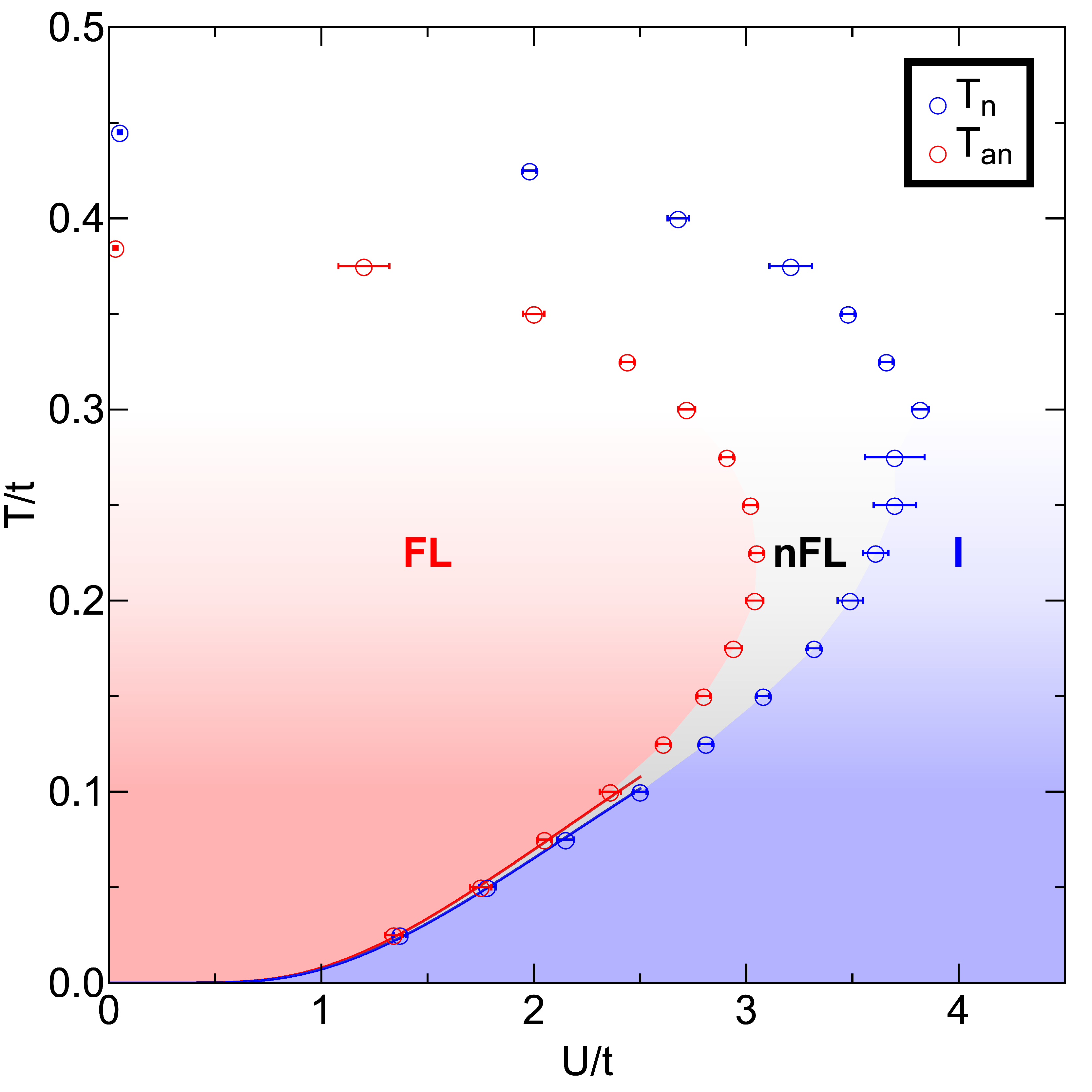}
\caption{\label{fig:phasediagram} Crossover lines between the Fermi-liquid (FL, red), non-Fermi-liquid (nFL, gray), and quasi-AFM insulator (I, blue) regimes of the half-filled 2D Hubbard model (\ref{H}) on the square lattice in the $U$-$T$ plane obtained by $\Sigma$DDMC. The solid lines fit data by the functions $T_{\text{an}}=a\exp(-b/\sqrt{U})$ and $T_{\text{n}}=a'\exp(-b'/\sqrt{U})$ with $\{a,b,a',b'\}=\{6.99,6.51,4.7,6.08\}$. It follows that below $U = 2.5t$ the low-temperature physics is of the mean-field character, while beyond $U=4t$ the low-temperature behavior is expected to resemble that of the Heisenberg model.}
\end{figure}

Our main result is summarized in Fig.~\ref{fig:phasediagram}. At sufficiently low temperature,
$T \lesssim 0.25t$, we observe the
crossover from the low-$U$ metallic FL state with a well-defined FS
(the red region) to the quasi-AFM insulator with
temperature-activated quasiparticles (the blue region).
In between, there is a transitional nFL regime (the gray region) bound by two lines $T_{\text{an}}(U)$ (red points) and $T_{\text{n}}(U)$ (blue points): as $T$ is lowered, the quasiparticle gap continuously develops
along the FS, first appearing at the antinodal point $\mathbf{k}_{\text{an}}=(\pi, 0)$ at $T_{\text{an}}(U)$
and proliferating to the nodal point $\mathbf{k}_{\text{n}}=(\pi/2, \pi/2)$ at $T_{\text{n}}(U)$.
Thus, the crossover involves a regime with a partially gapped FS and damped gapless quasiparticles
elsewhere on the FS, the so-called pseudogap driven by extending AFM fluctuations~\cite{Tremblay2006review, Jin2011, senechal2004pseudogap, macridin2006pseudogap, Jin2011, gunnarsson2015diagnostics, wu2017, wu2018}, similar to that found at the MIT for the model~(\ref{H}) in eight-site cluster-DMFT~\cite{Werner098site}. It extends over an appreciable range of parameters at larger
$U$ (or $T$). When quasiparticle properties could be meaningfully defined ($T \lesssim 0.25t$), we find that already at $U \gtrsim 4t$ the FL is lost and the self-energy reveals a charge gap. This value is significantly smaller than the critical $U_c \sim 5-6t$ found for the MIT in small-cluster DMFT results \cite{park:2008, Werner098site, gull:2013}. This leaves no room for the MIT in the Hubbard model (\ref{H}) without additional frustration of antiferromagnetic correlations---the FL-quasi-AFM crossover destroys the FL before it can undergo a first-order transition everywhere where the FL can be defined.
$U = 4t$ is the upper bound on the interaction strength beyond which the low-$T$ behavior
is not qualitatively different from that of the Heisenberg model.  As the crossover is expected to become increasingly mean-fieldlike at smaller $U <2.5t$, driven by magnetic correlations with large $\xi$, it is rather instructive that in this regime the crossover temperatures $T_{\text{an}}(U)$ and $T_{\text{n}}(U)$ approximately coincide and both are captured by the mean-field N\'{e}el temperature ansatz $a\exp(-b/\sqrt{U})$ with empirical parameters that agree with estimates found in Ref.~\cite{vsimkovic2017magnetic}.

We verify our results in Fig.~\ref{fig:phasediagram} against large-scale dynamical cluster approximation (DCA) calculations at higher temperatures.
DCA is a nonperturbative momentum space variant of cluster DMFT with which we utilize an auxiliary-field cluster impurity solver~\cite{ hettler:2000,jarrell:2001,Maier05}. Results for cluster sizes up to $144$ sites reveal
very slow convergence of the self-energy with cluster size. We further illustrate the significance of finite-size errors by comparing $\Sigma$DDMC results in the TDL to $\Sigma$DDMC calculations on finite lattices.

%%%%%%%%%%%%%%
%\section{Model and Methods}

In the FL theory, the quasiparticle residue at the chemical potential is a positive number less than unity.
It is defined through $\lim\limits_{\omega\to 0} \left[\partial \rm{Re}\Sigma_\mathbf{k} (\omega)/ \partial \omega \right]$ (with $\Sigma_\mathbf{k} (\omega)$ the self-energy at the momentum $\mathbf{k}$ and real frequency $\omega$). On the Matsubara (imaginary-frequency) axis, in the low-temperature limit, this is equivalent through a Wick rotation to $\lim\limits_{i\omega_n\to 0} \left[ \partial \rm{Im}\Sigma_\mathbf{k}(i\omega_n) / \partial \omega_n \right]$.
In contrast, $\Sigma_\mathbf{k} (\omega)$ in an insulator exhibits a pole at $\omega=0$.
At sufficiently low temperature, when discrete values of $\omega_n=2\pi T(n+1/2)$ remain closely spaced
to perform the limit $\omega_n \to 0$, this qualitative difference provides a metric to define
the state: if $\rm{Im}\Sigma_\mathbf{k} (i\omega_0) >\rm{Im}\Sigma_\mathbf{k} (i\omega_1)$ for all $\mathbf{k}$ on the FS the state is metallic (FL-like, referred to as FL below); when the reverse is true, $\rm{Im}\Sigma_\mathbf{k} (i\omega_0) <\rm{Im}\Sigma_\mathbf{k} (i\omega_1)$ on the whole FS, the system exhibits an insulating behavior by opening a quasigap at finite temperature.
Throughout we use a shorthand notation, $\Delta \Sigma_\mathbf{k} = \rm{Im}\Sigma_\mathbf{k} (i\omega_0) -\rm{Im}\Sigma_\mathbf{k} (i\omega_1)$, positive (negative) values of which imply FL (insulator) states.
This characterization loses its meaning in a thermal state when the first frequency $\omega_1 = 3 \pi T$
is of the order of the Fermi energy.

%%%%%%%%%%%%%%
\begin{figure}
\centering
\includegraphics[width=\linewidth]{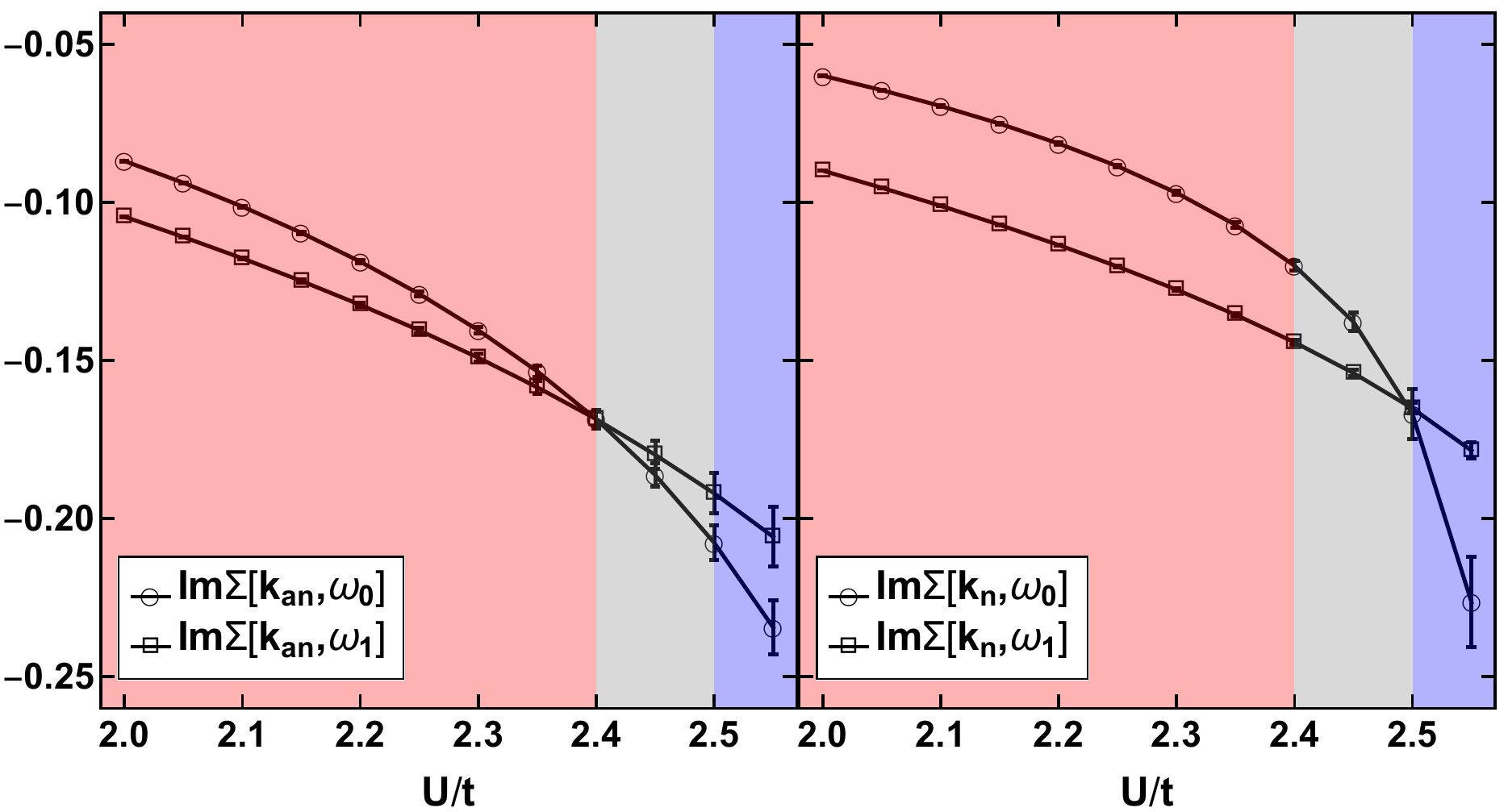}
\caption{\label{fig:freqdep}
Illustration of the FL-to-insulator crossover at $T/t=0.1$: evolution of the self-energy (imaginary part) at the two lowest Matsubara frequencies $\omega_0$ and $\omega_1$ at the momentum $\mathbf{k}_{\text{an}}=(\pi, 0)$ (left panel) and $\mathbf{k}_{\text{n}}=(\pi/2, \pi/2)$ (right panel) with increasing $U$. Colors correspond to FL, nFL, and quasi-AFM insulator regions of Fig.~\ref{fig:phasediagram}}
\end{figure}

$\Sigma$DDMC performs a numerically exact evaluation of the coefficients $a_n$ in the Taylor-series
expansion of $\Sigma$ in powers of $U$---which at half filling takes on the form
 \begin{equation}
\Sigma_{\mathbf{k} \sigma}(i\omega, T, \mu=U/2, U) = \sum_{n=1}^{\infty} a_{2n, \mathbf{k}\sigma}(i\omega, T) \, U^{2n}\,  \label{eqn:Sigma_series}
\end{equation}
---up to a maximal order $n_*$, the truncation of the series being the only approximation. 
In general, reconstructing $\Sigma_{\mathbf{k} \sigma}(i\omega)$ from its series is a problem by itself \cite{simkovic2017determinant}. In the regime of interest, however, the crossover lines are
always within the series convergence radius, and, in principle, the result can be obtained by taking the sum
up to a sufficiently high $n_*$ to ensure that the truncation error is negligible compared to the
statistical error. We are able to compute $\{a_{n}\}$ with statistical errors $\lesssim 10\%$ up to order $n_*=12$ for temperatures $T\geq 0.1$ and up to order $n_*=10$ for $0.025 \leq T\leq 0.1$. We further extrapolate the series by using the standard Dlog-Pad\'{e}-type approximants \cite{baker1961application,hunter1979methods},
and verify that the systematic error of the extrapolation procedure is small compared to the statistical error (see the Supplemental Material \cite{Supplemental}).

%%%%%%%%%%%%%%
%\section{Results}

Figure~\ref{fig:freqdep} illustrates how the crossover diagram, Fig.~\ref{fig:phasediagram}, was obtained. It shows the variation with $U$ of $\rm{Im} \Sigma(i\omega_0)$ and $\rm{Im} \Sigma(i\omega_1)$ at the antinodal and nodal points
($\rm{Re} \Sigma =0$ at the FS).
At $U=2t$ the values at $\omega_0$ are higher than at $\omega_1$, which is typical for a FL.
As $U$ is increased, $\Delta \Sigma_{\mathrm{k}}$ shows a trend toward nFL behavior by first changing its sign at $\mathbf{k}_\text{an}$ (we take it as the onset of the nFL behavior).
Following our measure, we mark the region of $U$ and $T$ where $\Delta \Sigma_{\mathbf{k}} >0$
for all momenta on the FS as FL (red shading). Similarly, the insulating region (blue shading)
corresponds to $\Delta \Sigma_{\mathbf{k}} <0$ for all $\mathbf{k}$ on the FS.
The nFL pseudogap regime (gray) falls in between the two: it has $\Delta \Sigma_{\mathbf{k}} >0$
at some momenta on the FS and $\Delta \Sigma_{\mathbf{k}} <0$ at others.
Correspondingly, the nFL state is bounded by the temperature scales $T_{\text{an}}(U)$ and $T_{\text{n}}(U)$ where
$\Delta \Sigma_{\mathbf{k}} =0$ at momenta $\mathbf{k}_\text{an}$ and $\mathbf{k}_\text{n}$, respectively. As the gap proliferates along the FS between $T_{\text{an}}(U)$ and $T_{\text{n}}(U)$, we expect that the heat capacity $C_V$ to $T$ ratio decreases before growing due to the AFM quasiorder near the insulator boundary and eventually reaching the asymptotic $C_V/T \propto T$ law due to spin waves deep in the insulating regime.

At small $U$, both crossover temperatures scale exponentially according to the BCS solution for the mean-field
AFM transition, $T_{\text{an}}=a\exp(-b/\sqrt{U})$ and $T_{\text{n}}=a'\exp(-b'/\sqrt{U})$, with the fit parameters
$a=6.99$, $b=6.51$, $a'=4.7$, $b'=6.08$ (see Fig.~\ref{fig:phasediagram}), suggesting the crossover is being
driven by extended AFM correlations. 

At high temperature $T \gtrsim 0.25t$, the data points corresponding to $\Delta \Sigma_{\mathbf{k}_{\text{an},\text{n}}} =0$ lose their meaning as boundaries between FL, nFL, and insulator regimes (white region in Fig.~\ref{fig:phasediagram}).
Given that $\rm{Im}\Sigma (i\omega_n)$ is a negative-valued function approaching zero at large frequencies,
it is clear that by increasing $T$ in the FL regime $\Delta \Sigma $ will change the sign, marking a crossover
to the thermal gas, not the nFL, state. Nonetheless, we plot the results for
$\Delta \Sigma_{\mathbf{k}_{\text{an},\text{n}}} =0$ at high temperature as TDL benchmarks
for other numerical techniques.

\begin{figure}
\centering
\includegraphics[width=\linewidth]{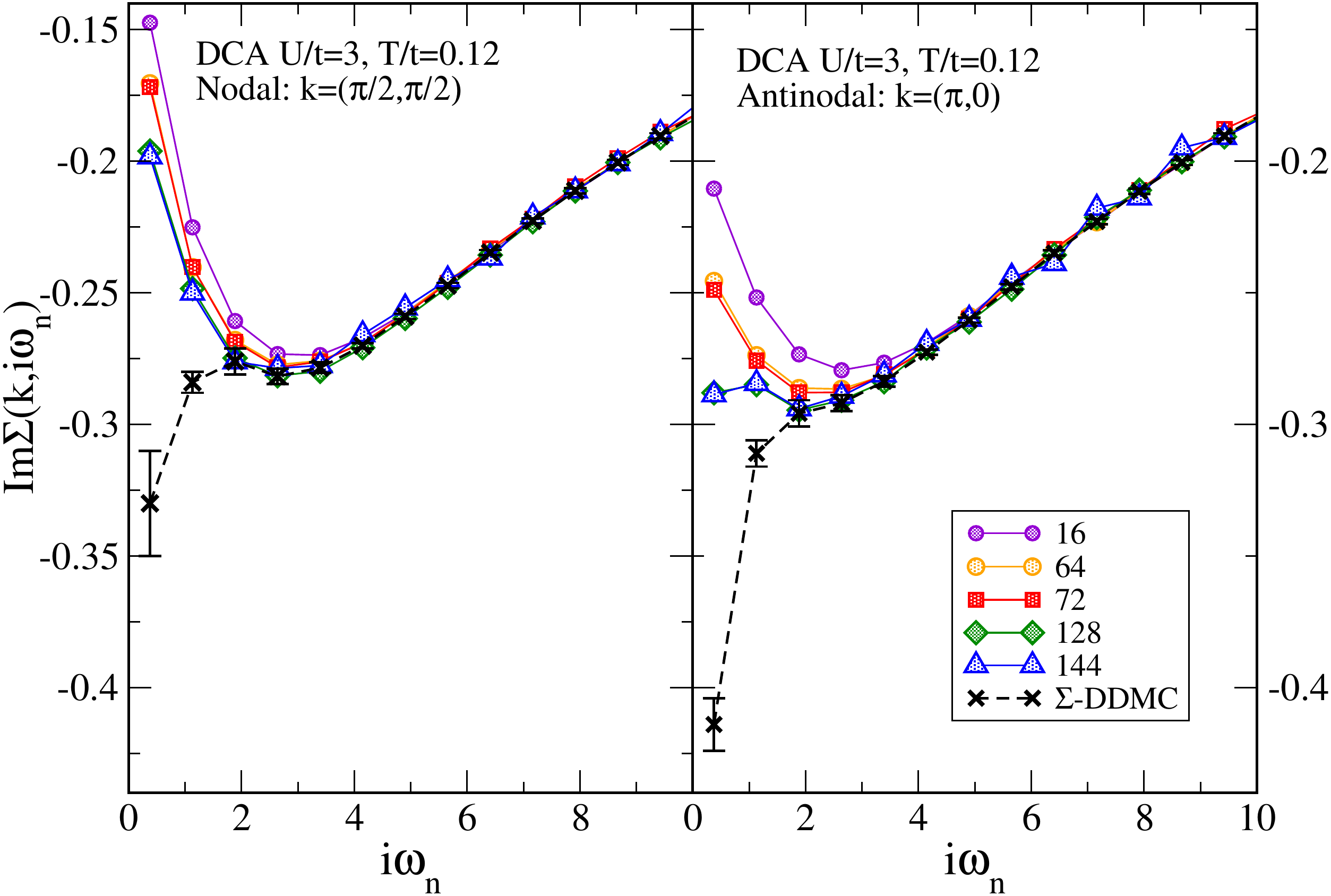}
\caption{\label{fig:dcavsw}  Imaginary part of the self-energy obtained by DCA and compared to $\Sigma$DDMC at $\mathbf{k}_{\text{n}}$ (left) and $\mathbf{k}_{\text{an}}$ (right) in the insulating regime, $U/t=3$, $T/t=0.12$ (cluster sizes are $N_c=16,64,72,128,144$). }
\end{figure}

Providing the controlled extrapolation of finite-system numerical results to the TDL has been long recognized as important. Thus far it has been accomplished, in particular, by large-system-size studies of the 3D Hubbard model near the AFM N\'{e}el transition~\cite{kent2005neel, fuchs:2011, kozik2013neel}. In 2D, the extremely slow finite-size scaling due to the exponential growth of the AFM correlation length near the crossover makes Fig.~\ref{fig:phasediagram} challenging to reproduce by finite-size methods, even if they do not suffer from the fermionic sign problem at half filling (cf. DQMC results for this problem in Ref.~\cite{schafer:2015}). In the Supplemental Material~\cite{Supplemental}, we further illustrate the difficulty of obtaining $T_{\text{an}}(U)$, $T_{\text{n}}(U)$ through extrapolation to the TDL by the example of $\Sigma$DDMC calculations on finite-size lattices of dimensions $L \times L$, where any value of $L$ is accessible at the same computational cost.

To verify our results by an independent method we resort to the DCA,
which produces unbiased results after extrapolation to the TDL. At low temperatures the extrapolation is extremely challenging. 
Figure~\ref{fig:dcavsw} shows DCA results for $\rm{Im}\Sigma_\mathbf{k}(i\omega_n)$ at $U=3t$, $T=0.12t$, and various cluster
sizes $16 \leq N_c \leq 144$. At $\mathbf{k}=\mathbf{k}_\text{n}$ (left panel), the results show the FL behavior for all accessible system sizes, while the $\Sigma$DDMC results are insulatorlike. At $\mathbf{k}=\mathbf{k}_\text{an}$ (right panel), the character changes from FL to insulating as a function of $N_c$ when cluster sizes exceed $100$, reaching qualitative agreement with the $\Sigma$DDMC data. Bigger clusters are required for a quantitatively accurate extrapolation. Note that if the data for only $N_c \leq 72$ were available, one would be led to conclude that the state
at $U=3t$, $T=0.12t$ is a FL. [Similarly, in the DQMC study of Ref.~\cite{schafer:2015}, small-system-size data in what is actually a FL regime show insulatorlike behavior, resulting in a large error bar of the extrapolation to the TDL.] It is not surprising then that past work limited to substantially smaller cluster sizes~\cite{park:2008, Werner098site, gull:2013} observed the MIT at $U_c>5t$, while the TDL system is, in fact,
already insulating at smaller $U$. In general, finite-size effects are less severe at higher $T$, and a controlled extrapolation of DCA data to the TDL is feasible: extrapolated DCA results for the crossover temperatures are in quantitative agreement with $\Sigma$DDMC for $T \gtrsim 0.2t$ (see the Supplemental Material~\cite{Supplemental}).

\textbf{\begin{figure}
\centering
\includegraphics[width=\linewidth]{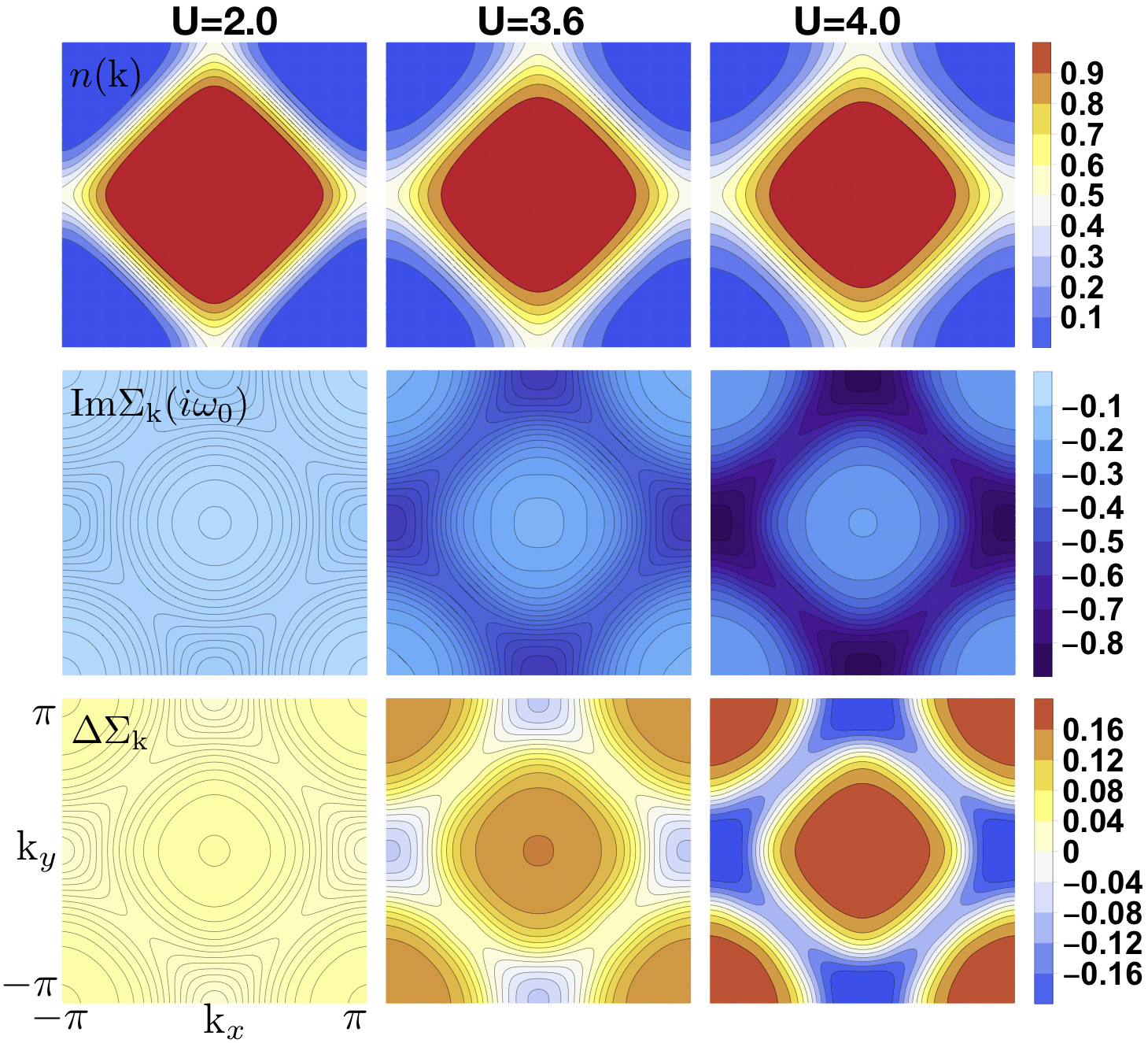}
\caption{\label{fig:BZ_map} (Quasi)momentum distribution $n(\mathbf{k})$ (top row), $\rm{Im}\Sigma_\mathbf{k}(i\omega_0)$ (middle row), and $\Delta \Sigma_\mathbf{k}$ (bottom row) in the Brillouin zone at $T=0.2t$ and different values of $U$ (columns).   }
\end{figure}}

Arbitrary momentum resolution of $\Sigma$DDMC allows us to observe the structure correlations in the momentum space. Figure~\ref{fig:BZ_map} shows the Brillouin zone map of $\rm{Im} \Sigma_\mathbf{k}(\omega_0)$ and $\Delta \Sigma_\mathbf{k}$ and relates them to the measurable with ultracold atoms (quasi)momentum distribution $n(\mathbf{k})$ at the experimentally relevant $T=0.2t$ and three values of $U$ that are seen in Fig.~\ref{fig:phasediagram} to be: in the metallic ($U=2t$), nFL ($U=3.6t$), and insulator ($U=4t$) regimes, respectively. As $U$ is increased, $| \rm{Im} \Sigma_\mathbf{k}(\omega_0)|$ exhibits relatively small change away from the FS and grows substantially along the FS, with the largest value at $\mathbf{k}_\text{an}$ and the smallest within the FS at $\mathbf{k}_\text{n}$. At the same time, $\Delta \Sigma_\mathbf{k}$ remains positive for all $\mathbf{k}$ in the metallic regime, but islands of negative values appear around $\mathbf{k}_\text{an}$ at $U=3.6t$ [below $T_{\text{an}}(U)$]. At $U=4t$ [below $T_{\text{n}}(U)$], $\Delta \Sigma_\mathbf{k}$ is negative along the whole FS, with the lowest (most insulatorlike) values around $\mathbf{k}_\text{an}$. Interestingly, away from the FS, the change in $\Delta \Sigma_\mathbf{k}$ is opposite: it grows, becoming more FL-like, upon increasing $U$. The corresponding $n(\mathbf{k})$ shows gradual smearing of the step at the FS as $U$ increases and the system crosses from the FL to insulating regime. This smearing is more pronounced around $\mathbf{k}_\text{an}$: the diamond-shaped region of occupied states in the metallic regime shrinks and evolves toward a circle across the crossover, providing an observable signature of developing correlations.

%\section{Conclusions}

In conclusion, we have revealed the scenario of the metal-to-quasiantiferromagnetic-insulator crossover in the 2D Hubbard model (\ref{H})
qualitatively different from the MIT previously suggested for this system~\cite{park:2008, Werner098site, gull:2013} and in qualitative agreement with the recent D$\Gamma$A~\cite{schafer:2015, rohringer2016} and DQMC~\cite{schafer:2015} results. The crossover could not be captured by the small-cluster DMFT restricted to the paramagnetic solution, predicting the MIT instead. The insulating regime sets in at all values of $U$ due to extended AFM correlations that transform the system into the quasi-AFM after an intermediate nFL pseudogap regime. The quantitative shape of the crossover is different from that reported in Refs.~\cite{schafer:2015, rohringer2016}: it is described by the mean-field AFM transition at small $U$ and features a nFL regime that transforms to insulating behavior below $U \approx 4t$. All our results are obtained with controlled accuracy and
offer guidance for precision experiments with ultracold atoms in optical lattices, as well as unbiased numerical techniques, in the ongoing effort to describe the phase diagram of the Hubbard model. In particular, the most nontrivial correlated regime is realized for $2.5t < U < 4t$ and temperatures $T \lesssim 0.25t$. At weaker coupling, $U<2.5t$, the low-temperature behavior is governed by the mean-field BCS-type physics, while at $U > 4t$ the low-temperature state is expected to qualitatively resemble that of the Heisenberg model~\cite{aaram:spin_charge2019}. By continuity of the key mechanism, the long-range AFM correlations (quantified, e.g., in Ref.~\cite{vsimkovic2017magnetic}), this qualitative crossover picture is valid in a range of nonzero next-to-nearest-neighbor hopping $t'$ and doping $\delta$. The question of whether the conventional MIT scenario is realized at certain (large-enough) $t'$ requires further systematic investigation.

%\section{Acknowledgments}

\acknowledgements{We are grateful to Antoine Georges, Alexander Lichtenstein, Andy Millis, Thomas Sch\"{a}fer, and Alessandro Toschi for discussions of the results. F\v{S}, AJK, and EK are grateful to the Precision Many-Body Group at UMass Amherst, where a part of this work was carried out, for hospitality. This work was supported by the Simons Foundation as a part of the Simons Collaboration on the Many-Electron Problem, the EPSRC through Grant No. EP/P003052/1, the National Science Foundation under the grant DMR-1720465, and the MURI Program ``Advanced quantum materials - a new frontier for ultracold atoms" from AFOSR. JPFL was funded by NSERC and computational resources were provided by Compute Canada.}

\bibliographystyle{apsrev4-1}
\bibliography{refs.bib}

\end{document}

%\linenumbers
\title{Supplemental Material for ``Extended crossover from Fermi liquid to quasi-antiferromagnet in the half-filled 2D Hubbard model''}

\maketitle
%\listoftodos

%\section{Introduction}

\begin{wrapfigure}{R}{0.45\textwidth}
\centering
\includegraphics[width=\linewidth]{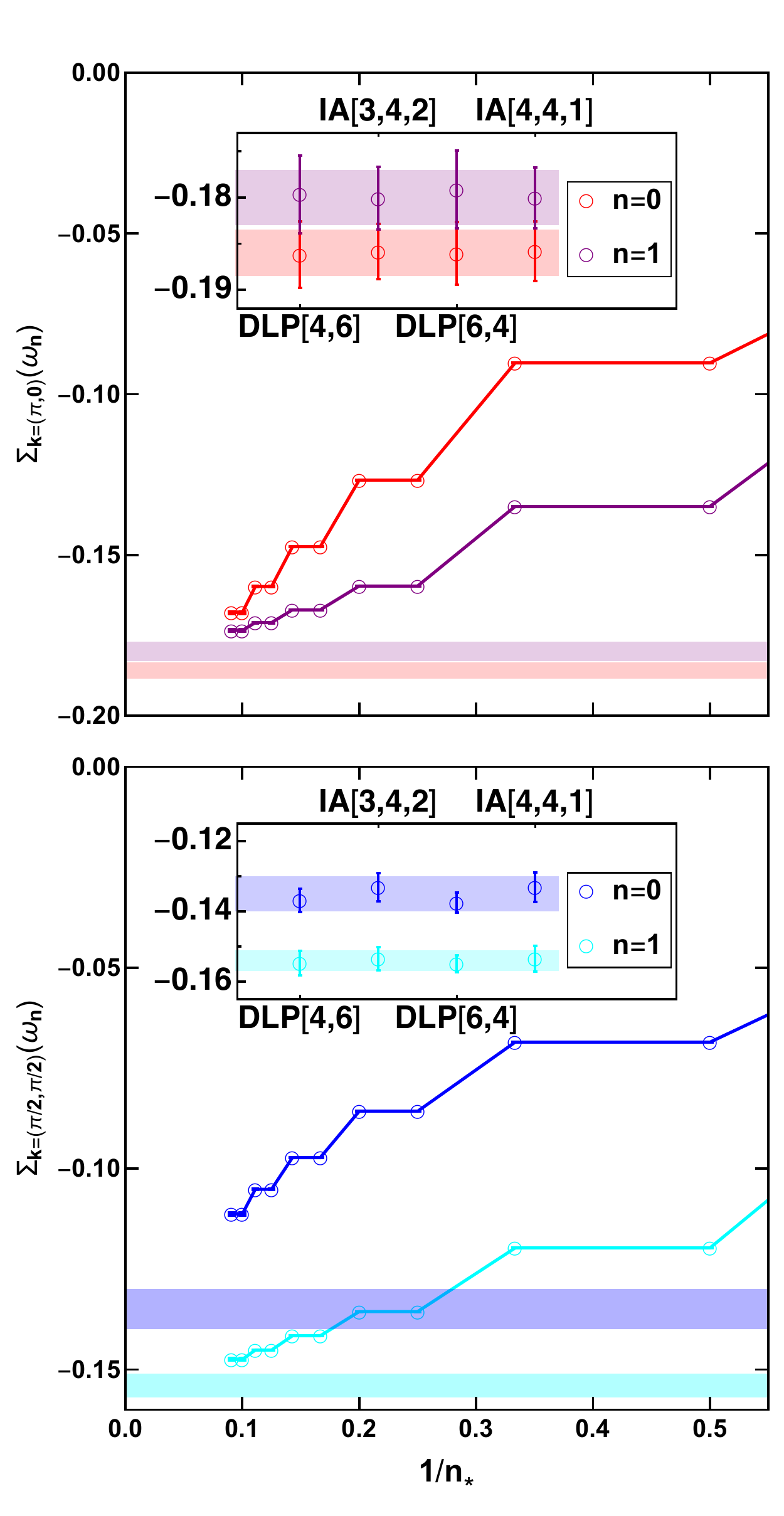}
\caption{\label{fig:order_by_order} Partial sum of the self-energy series $\Sigma_{\mathbf{k}}(i\omega)^{(n_*)}=\sum_{n=1}^{n_*} a_n(\mathbf{k}, \omega) U^n$ as a function of the inverse truncation order $1/n_*$ at the two lowest Matsubara frequencies $\omega_{0,1}$ and momenta $\mathbf{k}_\text{an}$ (upper panel) and $\mathbf{k}_\text{n}$ (lower panel) for $U=2.45t$ and $T=0.1t$. The insets illustrate the evaluation of $\Sigma = \lim_{n_* \to \infty} \Sigma^{(n_*)}$ using the IA and DLP methods, Eq.~(\ref{IA}), for various choices of $[L,M,N]$ and $[L,M]$, respectively (see text). The horizontal bands are the corresponding claimed results.}
\end{wrapfigure}

Fig.~\ref{fig:order_by_order} illustrates the behaviour of the self-energy series coefficients $a_n(\mathbf{k}, \omega)$ at the two lowest Matsubara frequencies $\omega_0$, $\omega_1$, and momenta $\mathbf{k}_\text{an}$ (upper panel) and $\mathbf{k}_\text{n}$ (lower panel): plotted is the dependence of the partial sum $\Sigma_{\mathbf{k}}(i\omega, U)^{(n_*)}=\sum_{n=1}^{n_*} a_n(\mathbf{k}, \omega) U^n$ on the inverse truncation order $1/n_*$ at the typical parameters $U=2.45t$ and $T=0.1t$. It is seen that the series converges but slowly, so that approximating the result for $\Sigma$ by $\Sigma^{(n_*)}$ is not sufficiently accurate. To evaluate $\Sigma(U) = \lim_{n_* \to \infty} \Sigma^{(n_*)}(U)$ we use the technique of integral approximants (IA)~\cite{hunter1979IA}, which associates the result with the function $g(x)$, $\Sigma=g(U)$, that has the same Taylor series as $\Sigma(U)$ up to the highest available order $n_*$ and satisfies the differential equation
\begin{equation}
Q_M(x) g'(x) + P_L(x) g(x) + R_N(x) = 0. \label{IA}
\end{equation}
Here $Q, P, R$ are polynomials of orders $M, L, N$, respectively, determined by substituting $g(x)=\sum_{n=0}^{n_*} a_n x^n$ in Eq.~(\ref{IA}) and solving it up to terms $\mathcal{O} (x^{M+N+L+2})$ with $M+N+L+2=n_*$. Eq.~(\ref{IA}) effectively continues the series for $g(x)$ from $n \leq n_*$ to infinite order and reconstructs the function behind it. The IA approach captures $\Sigma(U)$ of a general analytic structure with power-law singularities (see, e.g., Ref.~\cite{simkovic2017determinant} for details). The special case of Eq.~(\ref{IA}) with $R_N(x) \equiv 0$ is the so-called Dlog-Pad\'{e} (DLP) approach~\cite{baker1961Dlog}. To verify that the bias introduced by the extrapolation (\ref{IA}) is negligible, given the error bars of $\{a_n\}$ at hand, we repeat the procedure for several appropriate~\cite{hunter1979IA} choices of $[L,M,N]$ (for the general IA) and $[L,M]$ (for DLP) and make sure that the results agree within their respective error bars. The IA and DLP extrapolated results are shown in the insets in Fig.~\ref{fig:order_by_order}; the horizontal bands are the corresponding claimed values of $\Sigma$. 

\begin{wrapfigure}{R}{0.45\textwidth}
\centering
\includegraphics[width=\linewidth]{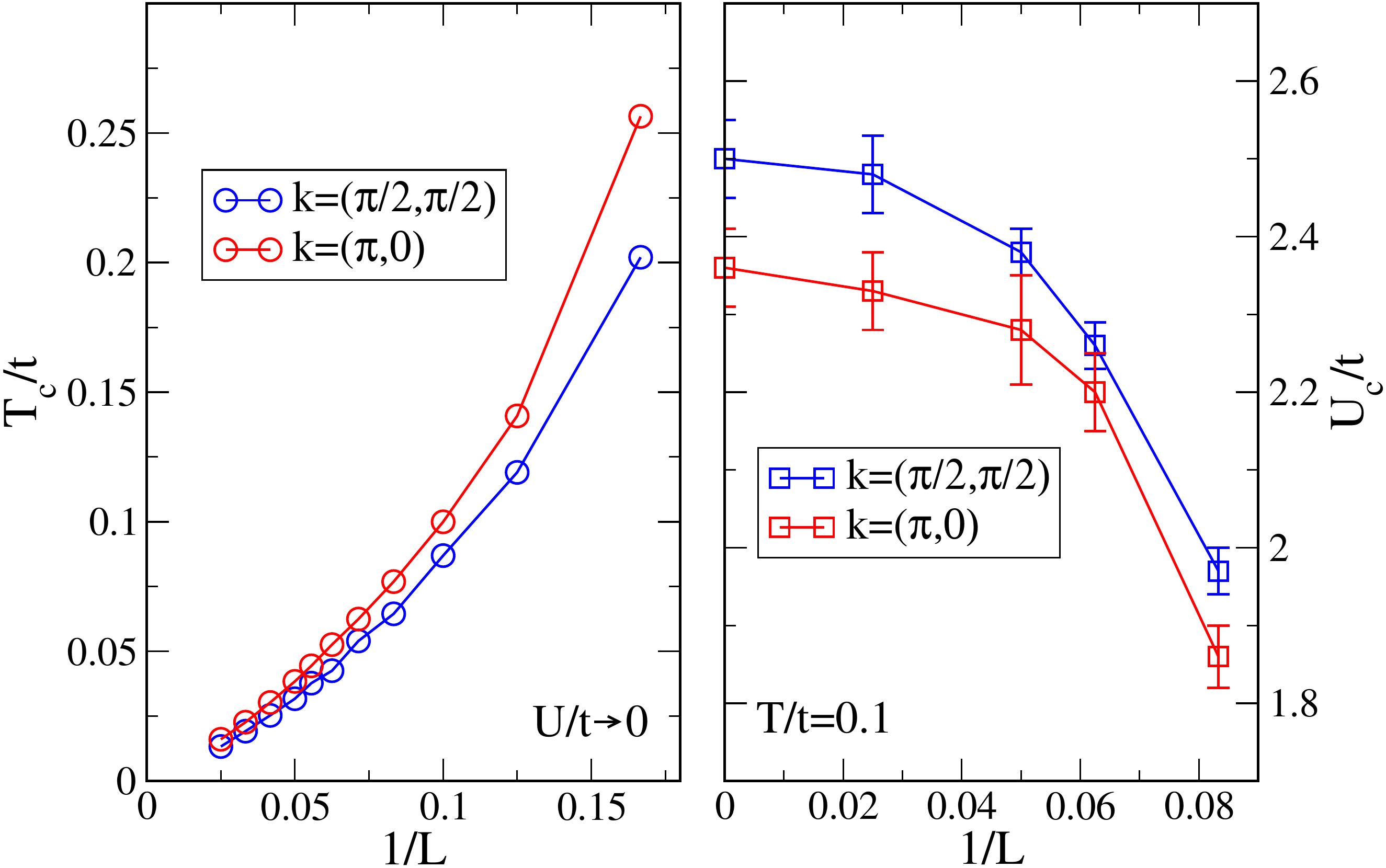}
\caption{\label{fig:weakcoupling}(Color online.) \emph{Left}:
Crossover temperatures in the weak-coupling limit $U/t\rightarrow 0$ as a function of inverse linear system size $1/L$. \emph{Right}:
Crossover interaction strengths $U_{\text{an}}$ and $U_{\text{n}}$ as a function of inverse linear system size $1/L$ for temperature $T/t=0.1$ obtained from $\Sigma$DDMC.}
\end{wrapfigure}

As an illustration and guidance for finite-size methods we show results for the crossover temperatures $T_{\text{an}}(U)$, $T_{\text{n}}(U)$ obtained by the same protocol
as in Fig.~1 of the main text but now using $\Sigma$DDMC for finite-size lattices of dimensions $L \times L$,
see Fig.~\ref{fig:weakcoupling}. The efficiency of $\Sigma$DDMC does not vary notably with $L$, so that any value
up to and including $L=\infty$ is accessible. The left panel addresses the weak-coupling regime, showing estimates of $T_{\text{an}}(U)$, $T_{\text{n}}(U)$,  found as solutions of $\Delta \Sigma_{\mathbf{k}_{\text{an}, \text{n}}}(U,T,L)=0$, for $U \to 0$ as functions of $1/L$, with $L$ ranging from $6$ up to $40$. Remarkably, the finite-size estimates span the whole range of temperatures from $T\sim 0.25t$ down to the correct TDL result $T_{\text{an}}(0)=T_{\text{n}}(0)=0$, making the extrapolation w.r.t. $L \to \infty$ from currently accessible system sizes ($L \sim 20$ in quantum Monte Carlo methods) problematic.
Similarly, the estimates of the corresponding crossover $U$ values at $T/t=0.1$ (right panel) show $\sim 30\%$ variation for $12 \leq L \leq 40$.

\begin{wrapfigure}{R}{0.45\textwidth}
\centering
\includegraphics[width=\linewidth]{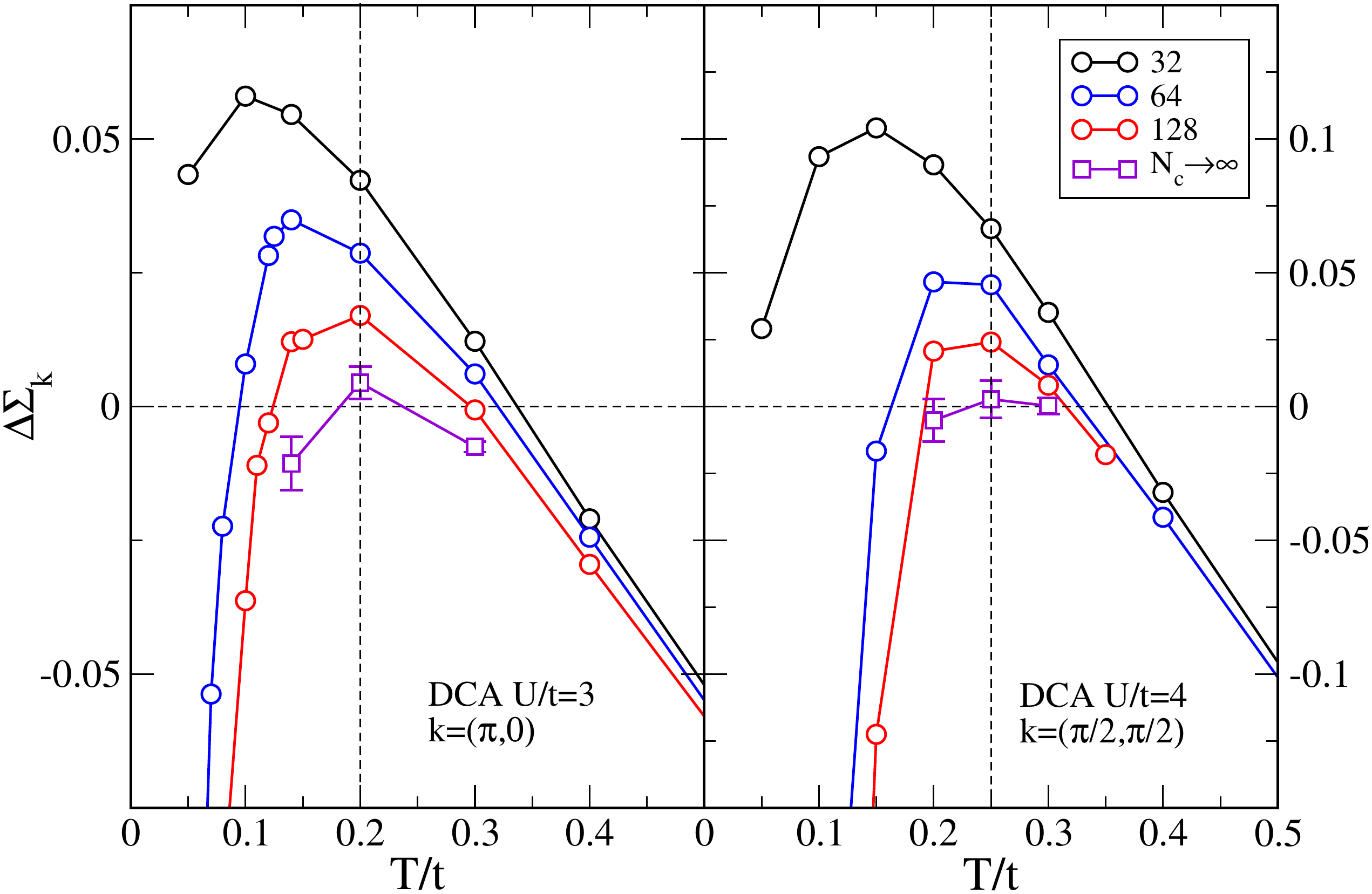}
\caption{\label{fig:U3vsT_AN} DCA results for $\Delta \Sigma_\mathbf{k}$ as a function of temperature at $U=3t$, $\mathbf{k}=\mathbf{k}_{\text{an}}$ (left) and $U=4t$, $\mathbf{k}=\mathbf{k}_{\text{n}}$ (right) for cluster sizes $N_c=32,64,128$. Also shown are several values in the TL obtained by extrapolation in $1/N_c \to 0$. The vertical dashed lines mark the crossover temperatures where $\Delta \Sigma_\mathbf{k}=0$ in $\Sigma$DDMC from Fig.~1 of the main text.   }
\end{wrapfigure}

In general, finite-size effects are less severe at higher $T$. At temperatures near and above $T=0.2t$ in the vicinity of the crossover, DCA results can be reliably extrapolated to the TDL, which we use to independently verify the existence of the crossover and the nFL pseudogap region. In Fig.~\ref{fig:U3vsT_AN} we show how the extrapolated DCA results reproduce the crossover temperatures predicted by $\Sigma$DDMC (vertical dashed lines) near $T=0.2t$, $U=3t$ for $\mathbf{k}=\mathbf{k}_\text{an}$ (left panel of Fig.~\ref{fig:U3vsT_AN}) and near $T=0.25t$, $U=4t$ for $\mathbf{k}=\mathbf{k}_\text{n}$ (right panel of Fig.~\ref{fig:U3vsT_AN}).
These points are in a region where the low-temperature FL/nFL/insulator and high temperature thermal states merge.
Both momenta for small clusters demonstrate the FL behaviour, $\Delta \Sigma_{\mathbf{k}}>0$, at low $T$ (not shown).
As the system size is increased above $N_c=32$ we see a dramatic dependence on $N_c$ at low $T$ with the
qualitative change from the FL to insulating behaviour.
To obtain results in the TDL we performed linear extrapolation of $\Delta \Sigma_{\mathbf{k}}$ in $1/N_c \to 0$. The extrapolated DCA answers reveal $\Delta \Sigma_{\mathbf{k}}=0$ at the values of $T$ in agreement with the respective $\Sigma$DDMC targets inferred from Fig.~1 of the main text and shown in Fig.~\ref{fig:U3vsT_AN}
by vertical dashed lines.

\bibliographystyle{apsrev4-1}
\bibliography{refs.bib}